\documentclass[aps,prl,twocolumn,showpacs]{revtex4}
\usepackage{graphicx}

\newcommand{\be}{\begin{equation}}
\newcommand{\ee}{\end{equation}}
\newcommand{\bea}{\begin{eqnarray}}
\newcommand{\eea}{\end{eqnarray}}

\newcommand{\cro}[1]{\left[#1\right]}

\newcommand{\avg}[1]{\langle{#1}\rangle}

\newcommand{\BE}{\begin{eqnarray}}
\newcommand{\EE}{\end{eqnarray}}
\newcommand{\BEn}{\begin{eqnarray*}}
\newcommand{\EEn}{\end{eqnarray*}}
\newcommand{\barr}{\begin{array}}
\newcommand{\earr}{\end{array}}

\newcommand{\bit}{\begin{itemize}}      
\newcommand{\eit}{\end{itemize}}
\newcommand{\bc}{\begin{center}}
\newcommand{\ec}{\end{center}}
\newcommand{\ben}{\begin{enumerate}}    
\newcommand{\een}{\end{enumerate}}

\begin{document}

\title{Bug propagation and debugging in asymmetric software structures}
\author{Damien Challet}
\affiliation{Theoretical Physics, Oxford University, 1--3 Keble Road, Oxford OX1 3NP, United Kingdom }
\author{Andrea Lombardoni}
\affiliation{Department of Computer Science, Eidgenossische Technische Hochschule, 8092 Z\"urich, Switzerland }

\begin{abstract}
  We address the issue of how software components are affected by the
  failure of one of them, and the inverse problem of locating the
  faulty component. Because of the functional form of the incoming
  link distribution of software dependence network, software is
  fragile with respect to the failure of a random single
  component. Locating a faulty component is easy if the failure only
  affects its nearest neighbors, while it is hard if it propagates
  further.
\end{abstract}
\maketitle

Large scale research on small-world networks began a few years ago
after the introduction by Watts and Strogatz of their famous
model~\cite{SW}. During the last few years, many real-life networks
turned out to be of small-world nature with a scale-free link
distribution~\cite{review}. The digital world seems particularly rich
in this type of network at all scales: wires in
computers~\cite{wires}, software function calls~\cite{Myers},
source-file dependencies~\cite{Moura}, software
modules~\cite{Sole,Potanin,Classes,Myers,Sole2}, Internet physical
network~\cite{Capocci}, and links between web
pages~\cite{Barabasi}. Are notably missing the networks between
software packages, which will be measured in the first part of this
paper.

Whereas previous work looked for reasonable explanations of why software
networks are scale-free~\cite{Myers,Sole2}, we address here bug
propagation and debugging in scale-free networks, a major issue that
has been neglected so far. We shall argue that software scale-free
networks provide a natural explanation of software fragility.

Scale-free networks in software were recently investigated in a game
and in the Java API (application programming interface)~\cite{Sole}.
The nodes were respectively the modules of the game (sound, graphics,
etc) and the objects of the standard Java API.  In both cases,
scale-free networks were discovered. 
 As noted in subsequent work~\cite{Myers,Potanin,Classes}, Ref.~\cite{Sole}
did not take into account the directed nature of these networks, which
are asymmetric. All these work focus on
microscopic software components, such as functions and objects. Here
we study the dependence between program packages in a Linux
distribution, an important structure which has not been
investigated yet, completing the hierarchy of scale-free networks
found in the digital world.  We then discuss the fragility of software
with respect to the failure of a single component and the difficulty
of debugging. As large networks are required for this study, we shall
also use function call networks of open-source projects.

\begin{figure}
\centerline{\includegraphics[width=0.4\textwidth]{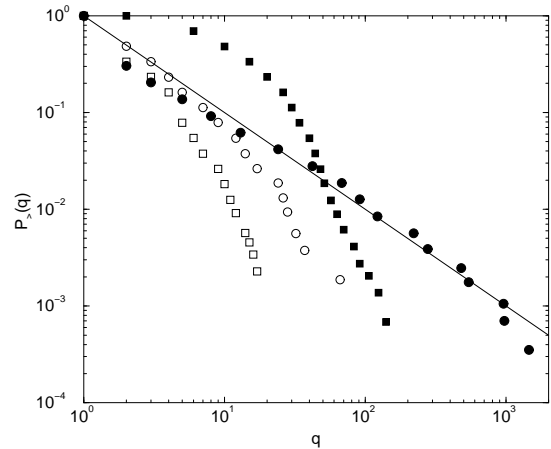}}
\caption{Cumulative distribution of incoming links (circles) and outgoing links
  (squares) between packages in a computer running RedHat 8.0. Empty
  and full symbols are obtained with {\tt rpm -q -\,-whatrequires} and
  {\tt rpm -q -\,-requires} respectively. The continuous line has a
  $-1$ slope.}
\label{inrh}
\end{figure}

Let us first study software components: a computer uses a collection
of software components that are linked through a network of
dependence. For instance a program that displays some text needs fonts
that can be provided by another component. In Linux distributions,
pieces of software are often provided as packages. As the name
indicates, a package is a collection of software components. The {\tt
rpm} command can be used to extract the network of package
dependences. More precisely, {\tt rpm -q -\,-whatrequires myprogram}
lists all the packages that needs {\tt myprogram}, making it easy to
build the package adjacency matrix.  One of us wrote a program called
{\tt rpmgraph} that produces a diagram of this
network~\cite{rpmgraph}. We studied an installation of RedHat 8.0 that
contained 1460 packages. The cumulative density $P_\ge(q)=P(q'\ge q)$
of the number of incoming links $q$ per package, is plotted in
Fig.~\ref{inrh}; we normalize $P_\ge(q)$ so that $P_\ge(1)=1$, which
amounts to leave out of $P_\ge(q)$ the nodes that are not needed by any
other node.  The distribution $P_\ge(q)$ has not enough points to be
fitted accurately with a power-law. The cumulative distribution of
outgoing links, $P_\ge(k)$ for $k>0$, is also plotted in
Fig. \ref{inrh}. The asymmetry between the outgoing and incoming link
distributions appears clearly.

 The {\tt rpm} command can give partial access to a more detailed network:
using the option {\tt -\,-requires} instead of {\tt -\,-whatrequires}
lists which components inside packages are needed by a given
package. In other words, a package is made up of a number of
sub-packages, each of them needing sub-packages of other packages. As
the rpm command can only be applied to a package, the sub-package
dependency network cannot be extracted, and we are left with the
distribution of the number of incoming links, and a coarse grained
distribution of outgoing links.  This provides however a much more
convincing evidence for the power-law nature of the incoming link
distribution: a fit over the whole dataset gives $P_\ge(q)\propto
q^{-\alpha+1}$ with $\alpha\simeq2.0$.

\begin{figure}
\centerline{\includegraphics[width=0.4\textwidth]{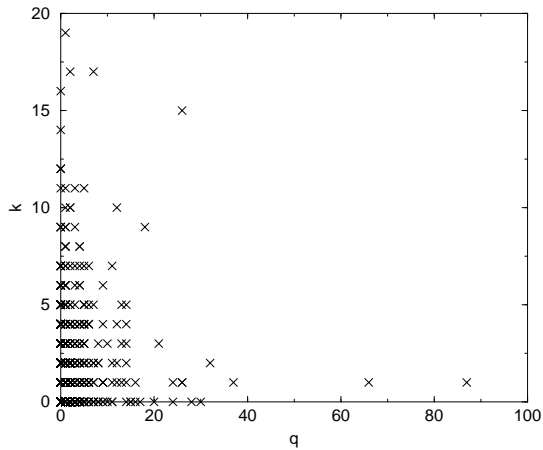}}
\caption{Number of outgoing links versus number of
  incoming links for the packages of RedHat 8.0}
\label{kq}
\end{figure}

 Fig.~\ref{kq} shows that the numbers of incoming and outgoing links of
 a given software package are generally correlated, a property also
 seen in links between functions in source code~\cite{Myers}. Simply
 put, this shows that some packages such as libraries provide a
 functionality to other programs. As we shall see below, this is one
 of the causes of software fragility.

\begin{figure}
\centerline{\includegraphics[width=0.4\textwidth]{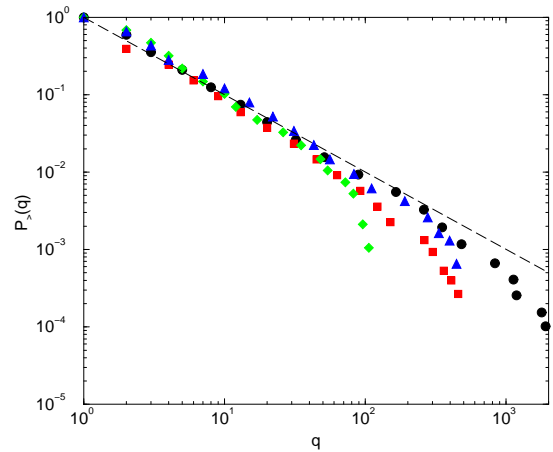}}
\caption{Cumulative distribution of incoming links in
  Linux (circles), Mozilla (squares), Apache (diamonds) and mySQL
  (triangles).  The dashed line is $1/q$.}
\label{in}
\end{figure}

In order to study bug propagation, we need better, more precise
 data. Therefore we will make use of function call networks: In the
 latter, functions are the nodes, and function calls are the links: in
 the following example written in C,
\begin{verbatim}
int f(int x){
  return 2*g(x);
}
\end{verbatim}

\begin{figure}
\centerline{\includegraphics[width=0.4\textwidth]{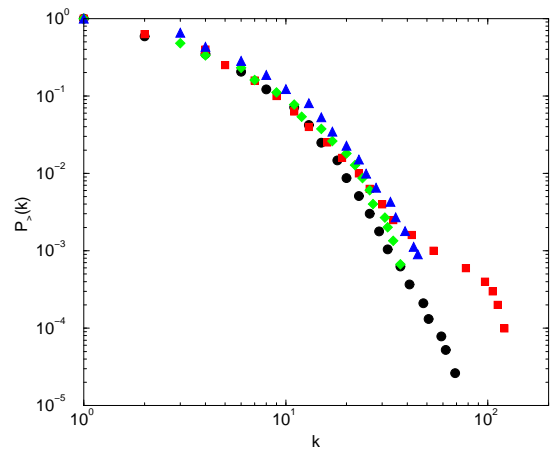}}
\caption{Cumulative distribution of outgoing links in
Linux (circles), Mozilla (squares), Apache (diamonds) and mySQL
(triangles).}
\label{out}
\end{figure}

{\tt f} calls {\tt g}, hence links to {\tt g}. Large open-source programs are ideal
candidates for investigation. Refs~\cite{Myers,Sole2} considered the
largest connected component. We are interested here in whole networks,
as we focus on bug dynamics and debugging. We studied Linux kernel
2.4.18, Mozilla Internet browser 1.3a, mySQL database 4.0.2, and
Apache web server 2.0.32. Extracting the function call network from a
source code written in C was done using simple scripts.  We excluded
C-keywords from the graph.  Fig.~\ref{in} reports that $P_\ge(q)$ is
also a power-law. It is noticeable that these data seem to suffer from
finite size effects similar to those seen for the package dependences,
the more data points, the closer to $2$ the exponent. 
We emphasize here that $\alpha=2$ is the value that marks
the border where the average number of times a piece of software is
used diverges when the size of the network goes to infinity. This is
possible in software because being reused does not cost anything to a
piece of software.  Therefore, the average number of programs that use
a given piece of software is free to diverge with the size of the
network.  The regularity of the incoming link distribution exponent
suggests some sort of universality: the source code and program
networks, that is, the micro- and mesoscopic levels respectively, have
roughly the same incoming link exponent. The latter is also very
close to exponents measured in macroscopic networks of links between
web pages and web sites, that is, of the phenomenology resulting from
the actual use of computers and programs. At all levels, being linked
is free for the nodes.

On the other hand, Fig \ref{out} shows the outgoing distribution,
which may have power-law parts, but it is impossible to assert it from
our data, because we have less than a decade of straight
line. Previous work fitted these distribution with a power-law
$P_\ge(k)\propto k^{-\beta+1}$ in the part that correspond to $k\le
10$ here and found exponents $\beta\simeq 2.4$. If this is the case,
there are strong cutoffs, much stronger than for the incoming link
distribution. On the other hand, it seems as reasonable to fit the
part $10<k<100$ with a power law, in which case a much larger exponent
(more than 4) is found. However, we only conclude from this graph that
the asymmetry between incoming and outgoing link distribution is
considerable, which is enough for our purpose.

 There are indeed special reasons for this asymmetry being more
 pronounced in software than in other structures.  As pointed
 out by Ref. \cite{Myers}, the asymmetry itself is due to software
 reuse: some pieces of software are designed to provide
 functionalities that other programs can use. In addition, writing a
 program, hence linking to previously written piece of software, is
 costly. As the number of dependences of a program is related to its
 complexity, the average number of outgoing links cannot diverge.
 Remarkably, the asymmetry of distributions is less pronounced for
 instance in links between web pages. We argue that linking to a web
 page can be almost free, in contrast to the amount of work needed to
 write software pieces, which needs a logical structure, hence the
 large asymmetry found here.

This leads us to bug propagation. Software is well known to be
fragile. As we shall argue in the following, this is due in part to
its structure. Assume that all the nodes but one, drawn at random, are
perfectly working. What is the consequence of this imperfection?
Software failures actually propagate on the dependence graph: if a
node (function or software package) is faulty, the nodes calling it
are likely to work less than optimally; by extension the nodes calling
a node that calls the faulty node will probably be affected, etc. This
also raise the question of how hard is it to locate the faulty node?

Interestingly, the failure can also propagate from a microscopic
software structure to a macroscopic one. For instance, a function
trying to access a memory address outside the allocated memory space
can crash the whole program to which it belongs. If it does, the
problem now lies at the level of software packages. If the operating
system has no memory protection, this causes a system crash. Then, if
other computers depend on the system that went down, they will also be
affected.

In this paper, we shall focus on a simpler problem by making
simplifying assumptions on the influence of a single bug. As many bugs
are not nearly as dangerous as illegal memory access, but (annoying)
imperfections, or faults, their influence is not as
dramatic. Therefore, we assume that the influence of a faulty node
is only determined by the dependence network to which it belongs. A node
is either working (contains no bug), faulty (contains a bug), or
affected by a bug.\footnote{This is somehow akin to virus
propagation \cite{epidemy} where a node is either susceptible,
infectious or resistant.}

\begin{figure}
\centerline{\includegraphics[width=0.4\textwidth]{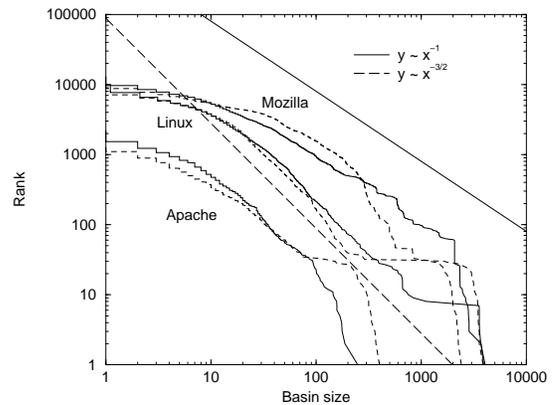}}
\caption{Basin size distribution of failure propagation on function
  call graphs. Continuous
  lines are for bug influence basins and dashed lines are for debugging
  basins.}
\label{basin}
\end{figure}

First we consider the simple optimistic case where only the nearest
neighbors are affected by a faulty node. The asymmetry of the
structure implies that the bug propagates to a typically large number of
nodes. On the other hand, once an incorrect behavior is detected,
locating the faulty node is easy. Therefore, in the most optimistic
case, software is fragile, but fixing it is relatively easy once an
anomaly is detected.

This view
is however too simplistic: as shown by the illegal memory example,
bugs do propagate further than their next neighbors. Let us be
pessimistic, and assume that they propagate as far as possible: if a
node is faulty, all the nodes that point to it directly or indirectly
are equally affected. In contrast to virus propagation, bug influence
is instantaneous. We are now left with the study of the properties of
influence basins. Of particular interest is the
influence basin size distribution $P(b)$ which can be computed by iterating the
graph matrix $G$~\cite{graphs}. The dependence of $i$ on $j$ is
denoted by $G_{i,j}=1$, ($G_{i,j}=0$ otherwise). Element $(G^n)_{i,j}$
contains the number of paths of length $n$ between $i$ and $j$, hence,
in order to compute the basin size distribution, one needs to compute
the $B=\sum_{k=0}^NG^k$ where $N$ is the number of nodes. If $i$
belongs to the influence basin of $j$, $B_{i,j}>0$. The size of
failure propagation basin of node $j$ is then simply given by
$b_j=\sum_i \textrm{sign}(B_{i,j})$.  Figure~\ref{basin} shows an
inverse Zipf plot of measured basin sizes: such a plot consists in
ranking the basins according to their sizes and plotting the rank $r$
versus $b$~\cite{Zipf}. This is equivalent to integrating: if
$P(b)\propto b^{-\gamma}$, $r\propto b^{-\gamma+1}$.\footnote{Usual Zipf plots
display $b$ versus the rank $r$, which is less intuitive, as the
apparent slope is $-1/(\gamma-1)$.}  The
exponents of the power-laws seem to be either $-2$ (Mozilla) or $-5/2$
(Linux); the exponent of Apache is unclear. A $-2$ exponent was also
obtained for the basins of Internet physical
network~\cite{basinInternet}.

\begin{figure}
\centerline{\includegraphics[width=0.4\textwidth]{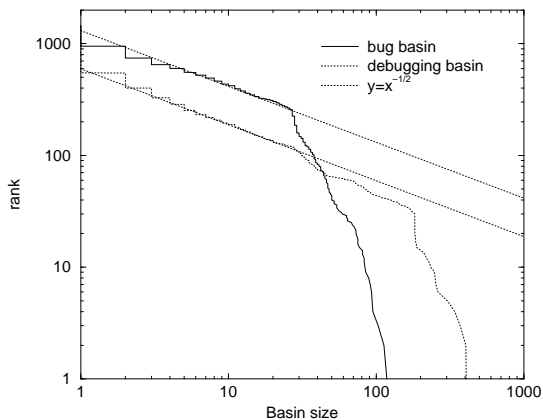}}
\caption{Basin size distribution of failure propagation on package
  dependence graph. The continuous line is for bug influence basins and dashed
  line is for debugging basins} 
\label{basins}
\end{figure}

One can also define a debugging basin: suppose that a piece of
software $i$ is affected by the failure of another program, but is not
buggy itself; what is the maximum number of pieces of software
$w_i=\sum_j B_{i,j}$ that are to be inspected in order to locate the
faulty node?  Given the asymmetry of incoming and outgoing link
distribution, one would naively expect that $P(b)$ and $P(w)$ differ
notably. This is clearly not the case: the debugging basin
distributions seem to follow closely their associated bug influence
basins distributions, and share roughly the same exponents (see
Fig~\ref{basin}), although $P(w)$ is not as smooth as $P(b)$, making
it difficult to fit it.

This similarity is also seen in the package
dependence network (fig \ref{basins}) where $P(b)$
and $P(w)$ have power-law parts both with same exponent $-3/2$; note
that our dataset is too small to allow being definitive. In addition,
the bug influence basis distribution as a early cut-off.

It is tempting to relate the similarity between the exponents of the
 two basin distributions to branching processes, which describe random
 tree growth.  Starting from a root node (generation 0), at time $t$
 each node $i$ of generation $t-1$ branches into a random number
 $r_i(t)$ of new nodes. The average number of new nodes $\avg{r}$ in
 the subtree is called the branching ratio. Of particular interest to
 us is the following property: if $\avg{r}=1$, the subtree
 (i.e. basin) size probability distribution of a randomly drawn node
 $P(b)\sim b^{-3/2}$. If $\avg{r}>1$ and $\avg{r^2}<\infty$, $P(b)\sim
 b^{-2}$ \cite{Paolo}; if the branching variance $\avg{r^2}$ is
 infinite, any exponent can be obtained \cite{PaoloPrivate}. These
 results do not apply directly to software structures, as the latter
 are not perfect trees. But what branching processes suggest is that
 the exponent of basin distributions is controlled by the branching
 ratio and variance. The branching ratio is nothing else than the
 average number of outgoing links $\avg{k}$ or incoming links
 $\avg{q}$ in the context of software structures, and both are equal.
 If the outgoing link distribution is a power-law and has an exponent
 small than $3$, both branching variances are infinite. At first, this
 provides an intuitive although incomplete explanation of why the
 basin distribution exponents are the same. Although the analogy is
 not perfect, it may be that there is also some sort of universality
 with respect to basin distributions in these networks, since the
 exponent found seem to be multiples of $1/2$.  This is an interesting
 open challenge. The question is whether larger exponents, hence more
 robust and easier-to-debug software, can be obtained at all. If the
 answer is negative, the fragility software and the difficulty of
 debugging are doomed not to be bounded in the worst case.

A still simple but more realistic model of bug propagation consists in
assuming that a node linking to a faulty or affected node is itself
affected with probability $p$. The rationale is the following: assume
that node $i$ calls node $j$. In the context of software packages, $p$
takes into account the fact that $j$, the faulty/affected node
contains sub-packages (see above) which are typically not all
defective/affected. Similarly, the sub-packages of $i$ do not all
link to a faulty sub-package of $j$. For instance, if there are $n$
sub-packages in both $i$ and $j$, and if there are $f$ faulty
sub-packages in $j$, and if every sub-package of $i$ has $l$ links
that point each to a randomly chosen sub-package of $j$, using
elementary combinatorics, one finds for $n-f>l$, \be
p=1-\cro{\frac{{n-f \choose l}}{{n \choose
l}}}^n=1-\cro{\frac{(n-f)!(n-l)!}{(n-f-l)!n!}}^n \ee where ${n-f
\choose l}/{n \choose l}$ is the probability that all the $l$ links
point to a working function. Assuming that $p$ is constant for all the
links in the network, one is left with a bond percolation problem for
directed graphs. It is known that if the exponent of the link
distribution is smaller than $3$ \cite{epidemy}, with probability 1 a
finite fraction of the network belongs to a percolation cluster, which
means that the influence of single bug is likely to be as large as if
$p=1$ for any value of $p$. Therefore, the picture drawn in the
previous paragraph and figs \ref{basin} and \ref{basins} still
applies. On the other hand, the short-tailed nature of the outgoing
link distribution implies that the basin of debugging depends on the
value of $p$: there is a critical value $p_c$ of $p$ such that for
$p_c<p$, debugging is easy, while debugging is hard if $p>p_c$.

Finally, another cause for software fragility comes from the peculiar role
played by libraries. As shown in fig \ref{kq}, software packages that
are meant to be reused are accordingly more often linked to. When a
program is installed or upgraded, it often happens that it needs an
updated version of some library. The dilemma, known as ``DLL hell'' in
Windows operating systems, is the following: if one
does not install the new version of the library, the new program is
likely not to work properly. If one updates the library, all the
programs that link to its old version are susceptible to be
broken. This provides a natural mechanism for progressive worsening of
operating system instability. There are two solutions: either one implements a
way of using several version of libraries at the same time, or one
systematically upgrades all the programs using the library in
question. Assuming that new versions of programs are available, the
first possibility applies mostly to commercial programs, because the
cost associated with upgrading expensive software may be very high; the
second solution is the way for instance Linux distributions work, but
leads sometimes to upgrading a  very large number of programs,
which is frowned up by the users. At any rate, one should not
underestimate the importance of this problem: not only the distribution
of incoming links implies that the average number of affected programs
diverges with the system size, but even worse, as shown by fig
\ref{kq}, libraries are characterized by an even larger number of
incoming links.

In conclusion, we argued that the fragility of software can be in part
attributed to its very structure, which unfortunately seems to
 arise naturally from optimization considerations.

D. C. thanks Paolo De Los Rios, Andrea Capocci and  Ginestra Bianconi for useful
discussions.





\end{document}